\documentstyle[aps,prl,preprint,tighten,floats]{revtex}   

\input{psfig}

\newcommand{\CP}{{\it CP}}
\newcommand{\Kz}{\mbox{$K^0$}}
\newcommand{\Kzbar}{\mbox{$\overline{K}$$^0$}}
\newcommand{\KS}{\mbox{$K_S$}}
\newcommand{\KL}{\mbox{$K_L$}}
\newcommand{\reepoe}{\mbox{$Re(\epsilon'/\epsilon)$}}
\newcommand{\ptsq}{\mbox{$p_T^2$}}
\newcommand{\EK}{\mbox{$E_K$}}
\newcommand{\delm}{\mbox{$\Delta m$}}
\newcommand{\taus}{\mbox{$\tau_S$}}

\begin{document}

\title{
{\flushright \normalsize
EFI-99-25 \\
\vspace*{1ex}
FERMILAB-Pub-99/150-E \\
}
\vspace*{5ex}
Observation of Direct {\it CP} Violation in $K_{S,L} \to \pi \pi$ Decays
}



\author{
A.~Alavi-Harati$^{12}$,
I.~F.~Albuquerque$^{10}$,
T.~Alexopoulos$^{12}$,
M.~Arenton$^{11}$,
K.~Arisaka$^2$,
S.~Averitte$^{10}$,
A.~R.~Barker$^5$,
L.~Bellantoni$^7$,
A.~Bellavance$^9$,
J.~Belz$^{10}$,
R.~Ben-David$^7$,
D.~R.~Bergman$^{10}$,
E.~Blucher$^4$, 
G.~J.~Bock$^7$,
C.~Bown$^4$, 
S.~Bright$^4$,
E.~Cheu$^1$,
S.~Childress$^7$,
R.~Coleman$^7$,
M.~D.~Corcoran$^9$,
G.~Corti$^{11}$, 
B.~Cox$^{11}$,
M.~B.~Crisler$^7$,
A.~R.~Erwin$^{12}$,
R.~Ford$^7$,
A.~Glazov$^4$, 
A.~Golossanov$^{11}$,
G.~Graham$^4$, 
J.~Graham$^4$,
K.~Hagan$^{11}$,
E.~Halkiadakis$^{10}$,
K.~Hanagaki$^8$,  
S.~Hidaka$^8$,
Y.~B.~Hsiung$^7$,
V.~Jejer$^{11}$,
J.~Jennings$^2$,
D.~A.~Jensen$^7$,
R.~Kessler$^4$,
H.~G.~E.~Kobrak$^{3}$,
J.~LaDue$^5$,
A.~Lath$^{10}$,
A.~Ledovskoy$^{11}$,
P.~L.~McBride$^7$,
A.~P.~McManus$^{11}$,
P.~Mikelsons$^5$,
E.~Monnier$^{4,}$\cite{monnier_on_leave},
T.~Nakaya$^7$,
U.~Nauenberg$^5$,
K.~S.~Nelson$^{11}$,
H.~Nguyen$^7$,
V.~O'Dell$^7$, 
M.~Pang$^7$, 
R.~Pordes$^7$,
V.~Prasad$^4$, 
C.~Qiao$^4$, 
B.~Quinn$^4$,
E.~J.~Ramberg$^7$, 
R.~E.~Ray$^7$,
A.~Roodman$^4$, 
M.~Sadamoto$^8$, 
S.~Schnetzer$^{10}$,
K.~Senyo$^8$, 
P.~Shanahan$^7$,
P.~S.~Shawhan$^{4}$, 
W.~Slater$^2$,
N.~Solomey$^4$,
S.~V.~Somalwar$^{10}$, 
R.~L.~Stone$^{10}$, 
I.~Suzuki$^8$,
E.~C.~Swallow$^{4,6}$,
R.~A.~Swanson$^{3}$,
S.~A.~Taegar$^1$,
R.~J.~Tesarek$^{10}$, 
G.~B.~Thomson$^{10}$,
P.~A.~Toale$^5$,
A.~Tripathi$^2$,
R.~Tschirhart$^7$, 
Y.~W.~Wah$^4$,
J.~Wang$^1$,
H.~B.~White$^7$, 
J.~Whitmore$^7$,
B.~Winstein$^4$, 
R.~Winston$^4$, 
J.-Y.~Wu$^5$,
T.~Yamanaka$^8$,
E.~D.~Zimmerman$^4$ \\
\vspace*{1ex}
(KTeV Collaboration)
\vspace*{1ex}
}

\address{
$^1$ University of Arizona, Tucson, Arizona 85721 \\
$^2$ University of California at Los Angeles, Los Angeles, California 90095 \\
$^{3}$ University of California at San Diego, La Jolla, California 92093 \\
$^4$ The Enrico Fermi Institute, The University of Chicago, 
Chicago, Illinois 60637 \\
$^5$ University of Colorado, Boulder, Colorado 80309 \\
$^6$ Elmhurst College, Elmhurst, Illinois 60126 \\
$^7$ Fermi National Accelerator Laboratory, Batavia, Illinois 60510 \\
$^8$ Osaka University, Toyonaka, Osaka 560 Japan \\
$^9$ Rice University, Houston, Texas 77005 \\
$^{10}$ Rutgers University, Piscataway, New Jersey 08855 \\
$^{11}$ The Department of Physics and Institute of Nuclear and 
Particle Physics, \\
University of Virginia, Charlottesville, Virginia 22901 \\
$^{12}$ University of Wisconsin, Madison, Wisconsin 53706 \\
}


\maketitle

\vspace*{5ex}

\begin{center}
Revised version submitted to {\it Physical Review Letters}, June 7, 1999
\end{center}

\clearpage

\begin{abstract}
We have compared the decay rates of \KL\ and \KS\ to $\pi^+ \pi^-$
and $\pi^0 \pi^0$ final states using a subset of the data from the
KTeV experiment (E832) at Fermilab.  We find that the direct-\CP-violation
parameter \reepoe\ is equal to $(28.0 \pm 3.0~\text{(stat)} \pm
2.8~\text{(syst)}) \times 10^{-4}$.  This result definitively
establishes the existence of \CP\ violation in a decay process.
\end{abstract}



The neutral $K$ meson system has been the subject of much study since
it was recognized that the two strangeness states (\Kz, \Kzbar) mix to
produce short- and long-lived kaons (\KS, \KL).  The
unexpected discovery of $\KL \to \pi \pi$ decays in 1964 \cite{ccft}
revealed that \CP\ (charge-parity) symmetry is violated by the weak
interaction, and it was soon understood that the dominant effect is an
asymmetry in the \Kz-\Kzbar\ mixing, parametrized by $\epsilon$.
Ever since, there has been great
interest in determining whether \CP\ violation also occurs in the
$K \to \pi\pi$
decay process itself, an effect referred to as ``direct'' \CP\
violation \cite{ww_review}
and parametrized by $\epsilon^\prime$.
This would contribute differently to the rates of $\KL \to
\pi^+\pi^-$ versus $\KL \to \pi^0\pi^0$ decays (relative to the corresponding
\KS\ decays), and thus would be
observable as a nonzero value of
%
%
%
\begin{displaymath}
\reepoe \approx \frac{1}{6} \left[ \frac{
\Gamma(\KL \to \pi^+\pi^-) / \Gamma(\KS \to \pi^+\pi^-)
}{
\Gamma(\KL \to \pi^0\pi^0) / \Gamma(\KS \to \pi^0\pi^0)
} - 1 \right].
\end{displaymath}
%

The standard Cabbibo-Kobayashi-Maskawa (CKM) model \cite{ckm} can
accommodate \CP\ violation in a natural way with a complex phase in
the quark mixing matrix.
The earliest standard-model calculations of \reepoe\
\cite{early_calc}, which gave values of order $10^{-3} \sim 10^{-2}$,
were done before the top quark mass was known and before the
importance of certain diagrams was appreciated.
Modern calculations depend
sensitively on input parameters and on the method used to estimate the
hadronic matrix elements.  Most recent estimates have tended toward
values near or below $10^{-3}$, for example $(4.6 \pm 3.0) \times 10^{-4}$
\cite{ciuchini} and $(8.5 \pm 5.9) \times 10^{-4}$ \cite{buras98};
however, one group has estimated a larger range of values,
$(17^{+14}_{-10}) \times 10^{-4}$ \cite{bertolini}.
Alternatively, a ``superweak'' interaction \cite{superweak} could
produce the observed \CP-violating mixing but would give $\reepoe =
0$.  Therefore, a nonzero value of \reepoe\ rules out
the possibility that a superweak interaction is the sole source of
\CP\ violation.

The two most precise past measurements of \reepoe\ are in only fair
agreement: the Fermilab E731 experiment reported $\reepoe = (7.4 \pm
5.9) \times 10^{-4}$ \cite{e731}, while CERN NA31 found $\reepoe = (23
\pm 6.5) \times 10^{-4}$ \cite{na31}.
%
%
Because of the importance of definitively
establishing the existence of direct \CP\ violation and determining
its magnitude, new experiments have been undertaken at Fermilab, CERN,
and Frascati to measure \reepoe\ with precisions of $(1 \sim 2) \times
10^{-4}$.
%


This Letter reports a new measurement of \reepoe\ using
23\% of the data collected by the KTeV experiment (E832) during the
1996-97 Fermilab fixed-target run.  The KTeV experiment was designed
to improve upon the previous generation of experiments and ultimately
to have the sensitivity to establish direct \CP\ violation in the
range of the smaller estimates in \cite{ciuchini} and \cite{buras98}.
The experimental technique is the same as in E731 \cite{e731prd}, and
differs from NA31 in two key ways.  First, it uses two kaon beams
from a single target to
enable the simultaneous collection of \KL\ and \KS\ decays in order to
be insensitive to the inevitable time variation of beam characteristics and
detector inefficiencies.  Second, it uses a precision magnetic
spectrometer to minimize backgrounds in the $\pi^+ \pi^-$ samples and
to allow {\it in situ} calibration of the calorimeter.  While the
method of producing the \KS\ beam (by passing a \KL\ beam through a
``regenerator'') is also the same as in E731, the KTeV regenerator is
made of scintillator and is fully instrumented to
reduce the scattered-kaon background to the coherently-regenerated signal.
A new beamline was constructed for KTeV with much cleaner beam
collimation and improved muon sweeping \cite{beam}.
Finally, the KTeV electromagnetic calorimeter has much higher
precision than the E731 calorimeter, permitting more accurate $\pi^0
\pi^0$ reconstruction and better background suppression.
%

%
\begin{figure}
\centering
\hspace*{0.0in}\psfig{figure=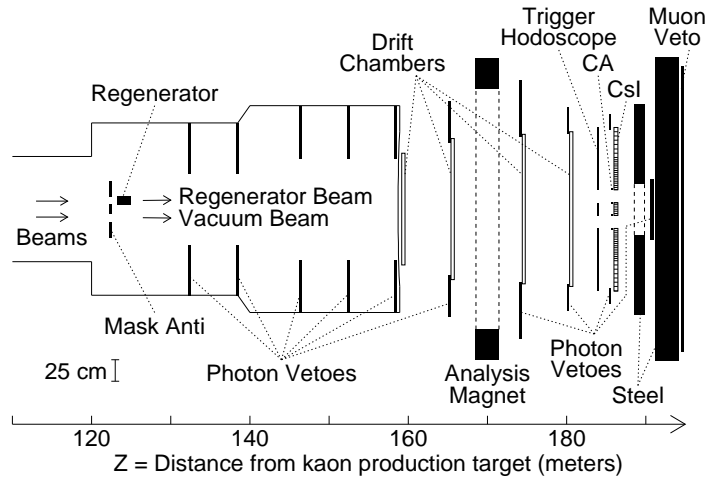,width=3.9in}
\caption{Plan view of the KTeV apparatus as configured to measure
  \reepoe.  The evacuated decay volume ends with a thin vacuum window at
  $Z=159$~m.  The label ``CsI'' indicates the electromagnetic
  calorimeter.}
\label{fig:detector}
\end{figure}

\begin{figure}
\centering
\hspace*{0.0in}\psfig{figure=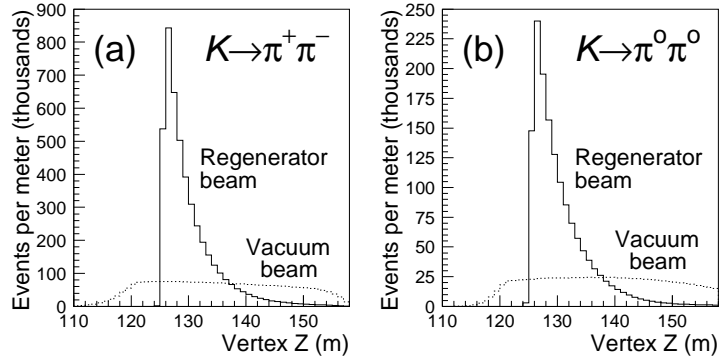,width=3.9in}
\caption{Decay vertex distributions for the (a) $K \to \pi^+\pi^-$ and (b)
$K \to \pi^0\pi^0$ decay modes, showing the difference between the
``regenerator'' (\KS) and ``vacuum'' (\KL) beams.}
\label{fig:zvacreg}
\end{figure}

Figure~\ref{fig:detector} shows the two beams (called ``regenerator''
and ``vacuum'') in the evacuated decay volume, with the main detector
elements located downstream.  The regenerator alternates sides 
between accelerator extractions
to minimize the effect of any left-right beam or detector
asymmetry.
A ``movable absorber,'' far upstream, attenuates the beam incident on
the regenerator.
To measure the double ratio of decay rates in the expression for
\reepoe, we must understand the {\em difference} between the
acceptances for \KS\ versus \KL\ decays to each $\pi\pi$ final state.
Triggering, reconstruction and event selection are done with identical criteria
for decays in either beam, so the principal difference between the
\KS\ and \KL\ data samples is in the decay vertex distributions, shown
in Fig.~\ref{fig:zvacreg} as a function of $Z$, the distance from the
kaon production target.
Therefore, the most crucial requirement for measuring \reepoe\ with
this technique is a precise understanding of the $Z$-dependence of
the detector acceptance \cite{na31train}.

In the regenerator beam, the beginning of the decay region is sharply
defined by a lead-scintillator module at the downstream end of the
regenerator.  In the vacuum beam, the acceptance for decays upstream
of $Z=122$~m is limited by the ``mask anti'' (MA), a lead-scintillator
counter with two square holes 50\% larger than the beams.

The KTeV spectrometer consists of four rectangular drift chambers,
each with two horizontal and two vertical planes of sense wires, and a
large dipole magnet which imparts a transverse momentum of
0.412~GeV$/c$.  The spaces between the drift chambers are filled with
helium to reduce scattering.  The drift chambers measure
horizontal and vertical track position 
with a typical resolution of 110~$\mu$m and 
momentum with a resolution of 0.4\% at the mean pion momentum of 36~GeV/c.
%

The electromagnetic calorimeter \cite{csical} consists of 3100 blocks
of pure cesium iodide (CsI) in a square array 1.9~m on a side and
0.5~m deep.  Two 15~cm square beam holes allow passage of
the neutral beams through the calorimeter.
%
%
The calorimeter was calibrated using $1.9 \times 10^8$
momentum-analyzed electrons from $\KL \to \pi e \nu$ decays collected
during normal running.
Individually tuned wrapping of the CsI blocks and the development of
very-low-noise readout electronics \cite{dpmt} have enabled the
calorimeter to achieve an average energy resolution of 0.7\% for
photons from $\pi^0\pi^0$ decays, which have a mean energy of 19~GeV.
%

%
The inner aperture for photons at the CsI is sharply defined by a
tungsten-scintillator ``collar anti'' (CA) counter around each beam
hole.  In addition, there are ten lead-scintillator ``photon veto''
counters to detect particles escaping from the decay volume or missing
the CsI, in order to suppress
$\KL \to 3\pi^0$ background in the $\pi^0 \pi^0$ samples.

The trigger system initiates detector readout based on synchronous
signals from a scintillator hodoscope located upstream of the
calorimeter (for $\pi^+\pi^-$) or on a fast analog energy sum from
the calorimeter (for $\pi^0\pi^0$).  To keep the trigger rate at a
manageable level, triggers are inhibited by fast veto signals from the
regenerator, the MA, a subset of the photon vetoes, and a downstream
hodoscope located behind 4~m of steel to detect muons \cite{deadtime}.
%
%
For $\pi^+\pi^-$, additional requirements are made on
the number and pattern of hits in the drift chambers.  For
$\pi^0\pi^0$, a hardware processor \cite{hcc} must find 4 or 5 ``clusters''
of energy in the calorimeter.  After readout, a CPU-based
``Level~3 filter'' reconstructs events and applies some loose kinematic
cuts to select $\pi^+ \pi^-$ and $\pi^0 \pi^0$ candidates.  Besides
these signal modes, large samples of $\KL \to \pi e \nu$, $\KL
\to \pi^+ \pi^- \pi^0$, and $\KL \to 3\pi^0$ decays are recorded for
detector calibration and acceptance studies.  In addition, an
``accidental'' trigger is formed, using scintillation counters near
the kaon production target, to randomly record the underlying activity
in the KTeV detector with the same instantaneous-intensity
distribution as the physics data.

The $\pi^0 \pi^0$ samples used for this analysis are from the data
collected in 1996, while the $\pi^+ \pi^-$ samples are from the first
18 days of data collected in 1997.  We decided not to use the $\pi^+
\pi^-$ data from 1996 because the Level~3 filter had a
22\% inefficiency arising from an unanticipated drift chamber effect
which sometimes delayed a hit by 20~ns or more.  The inefficiency was
nearly the same for both beams but would still have led to a large
systematic error on \reepoe.  The
Level~3 software was modified for the 1997 run to allow for this
effect, resulting in an inefficiency of less than 0.1\%.
%
%
%
Using $\pi^+ \pi^-$ and $\pi^0 \pi^0$ data from different running
periods does not significantly increase the systematic error on
\reepoe\ because the two modes use essentially independent detector
systems; detector inefficiencies and sources of deadtime cancel in the
\KS/\KL\ ratio for either mode independently.  The only direct effect
is a possible difference in the \KS/\KL\ flux ratio, which will be
discussed later.


For offline selection of $\pi^+ \pi^-$ candidates, each pion is
required to have a momentum of at least 8~GeV$/c$ and to deposit less
than 85\% of its energy in the calorimeter.  In order to cleanly
define the acceptance and to avoid topologies with poor reconstruction
efficiency, cuts are made on the distance from each pion to
the edges of the drift chambers, calorimeter, MA, and CA, and on the
separation distance between the two pions at the drift chambers and
calorimeter.  The $\pi^+ \pi^-$ invariant mass is required to be
between 488 and 508~$\text{MeV}/c^2$ (where the mean resolution is
approximately 1.6~$\text{MeV}/c^2$) and the square of the transverse
momentum of the $\pi^+\pi^-$ system relative to the initial kaon
trajectory, \ptsq, is required to be less than 250~MeV$^2/c^2$.

%
After applying various corrections to the raw calorimeter information,
$\pi^0 \pi^0$ candidates are reconstructed from four-photon events by
choosing the photon pairing combination which is most consistent with the
hypothesis of two $\pi^0$ decays at a common point, interpreted as the
kaon decay vertex.
Each photon is required to have an energy of at least 3~GeV and to be
at least 5~cm from the outer edge of the CsI and 7.5~cm from any other
photon.
The four-photon invariant mass is required to be between 490 and
505~$\text{MeV}/c^2$, where the mean mass resolution is approximately
1.5~$\text{MeV}/c^2$.
The initial kaon trajectory is unknown, so the only available indicator
of kaon scattering is the position of the energy centroid of the four
photons at the CsI.
This is used to calculate a ``ring number'', defined as four times the
square of the
larger normal distance (horizontal or vertical), in centimeters, from
the energy centroid to the center of the closer beam.  Its value is
required to be less than 110, which selects events with energy
centroid lying within a square region of area 110~cm$^2$ centered on
each beam.

In both the $\pi^+\pi^-$ and $\pi^0\pi^0$ analyses, cuts are made on
energy deposits in the MA, photon veto counters, and regenerator.  The
final samples consist of events with $110 < Z < 158$~m and $40 < \EK <
160$~GeV.


A detailed Monte Carlo (MC) simulation is used to determine the
detector acceptance for the $\pi\pi$ signal modes and to evaluate
backgrounds.
The simulation models kaon production and regeneration to
generate decays with the same energy and $Z$ distributions as the
data.
The decay products are traced through the KTeV detector, allowing for
electromagnetic interactions with beamline material and for pion decay.
The acceptance is largely determined by the geometry of the detector
and by geometric analysis cuts; however, to understand reconstruction
biases it is important to simulate the detector response accurately.
Energy deposits in the CsI blocks from photons, pions, and electrons are
based on the GEANT package \cite{geant}.  Drift chamber inefficiencies
and the delayed-hit effect are simulated using parametrizations and
position dependences measured from $\pi^+\pi^-$ data.
$\pi e \nu$ and $3\pi^0$ data samples are used to check or tune
various aspects of the detector geometry and simulation.
To reproduce possible
biases due to underlying activity in the detector,
an event from the accidental trigger is overlaid on top of each
simulated decay; the net effect on the measured value of
\reepoe\ is of order $10^{-4}$.  MC event samples are subjected
to the same reconstruction and selection criteria as the data samples.

\begin{figure}
\centering
\hspace*{0.0in}\psfig{figure=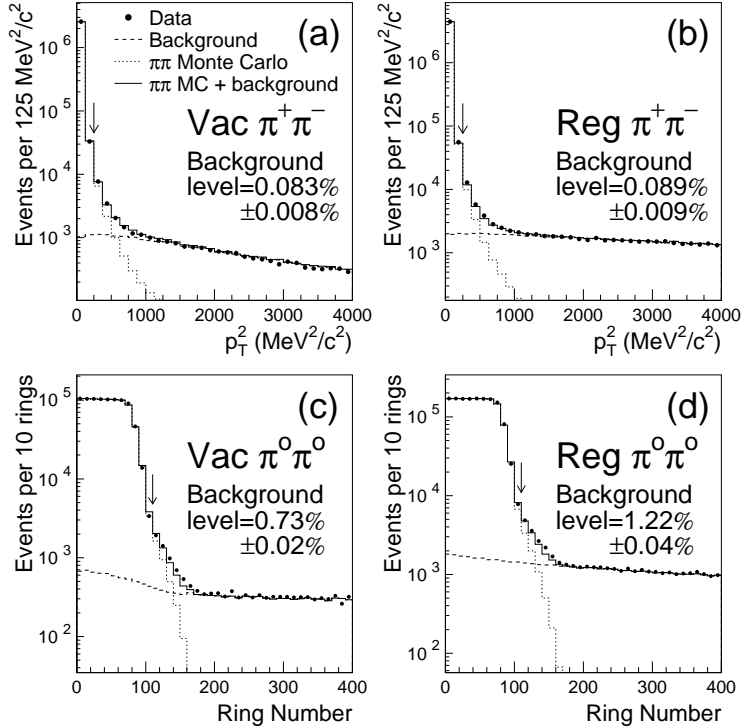,width=3.9in}
\caption{
Distributions of \ptsq\ for the $\pi^+\pi^-$ samples and ring number
for the $\pi^0\pi^0$ samples.
Total background levels and uncertainties (dominated by systematics)
are given for the samples passing the analysis cuts (arrows).
}
\label{fig:bkgplots}
\end{figure}

Background contributions to the $\pi^+\pi^-$ samples are determined by
using sidebands in the mass and \ptsq\ distributions to normalize MC
predictions from the various background processes.  Figure
\ref{fig:bkgplots} (a) and (b) show that the \ptsq\ distributions for
data are well described by the sum of coherent $\pi\pi$
MC and total background MC.
$\KL \to \pi e \nu$ and $\KL \to \pi \mu \nu$ decays, with the
electron or muon misidentified as a pion, contribute 0.069\% (0.003\%)
to the vacuum (regenerator) beam.  The dominant regenerator-beam
background (0.072\%) is from kaons which scatter in the regenerator
before decaying to $\pi^+\pi^-$.  Kaons which scatter in the final
beam-defining collimator contribute an additional 0.014\% to each
beam.
Data samples of $\pi^+\pi^-$ decays from kaons which scatter in the
regenerator or collimator
are used to tune physics-motivated scattering models incorporated into
the MC simulation.
%
%
%

%
The background levels are much larger for the $\pi^0 \pi^0$ samples
since the ring-number variable is not as effective as \ptsq\ at identifying
scattered kaons and cannot detect ``crossover'' scattering from the
regenerator into the vacuum beam.
Ring-number distributions are shown in
Fig.~\ref{fig:bkgplots} (c) and (d).  The upturn under the peak in
(c) is due to $\KL \to 3\pi^0$ decays with lost and/or overlapping
photons; it is determined, using mass sidebands, to contribute a
background of 0.27\% (0.01\%) to the vacuum (regenerator) beam.
A ring-number sideband (286-792) is used to normalize MC distributions
from kaons that scatter before decaying
to $\pi^0\pi^0$.  The vacuum (regenerator) beam background includes
0.30\% (1.07\%) from regenerator scattering and 0.16\% (0.14\%) from
collimator scattering.  Pairs of $\pi^0$'s produced by hadronic
interactions in the regenerator contribute an additional background
of 0.01\% in that beam.

After background subtraction, the net yields are 2,607,274
$\pi^+\pi^-$ in the vacuum beam, 4,515,928 $\pi^+\pi^-$ in the
regenerator beam, 862,254 $\pi^0\pi^0$ in the vacuum beam, and
1,433,923 $\pi^0\pi^0$ in the regenerator beam.


\reepoe\ is extracted from the background-subtracted data using a
fitting program which calculates decay vertex
distributions, properly treating regeneration and \KS-\KL\ interference
(including the residual \KS\ component in the vacuum beam at high energy).
The acceptance correction (as determined from MC) is applied, and the
resulting prediction for each decay mode
is integrated over $Z$ and compared to the data
in 10-GeV bins of kaon energy.  {\it CPT} symmetry is assumed, and the
values of the \KL-\KS\ mass difference (\delm) and \KS\ lifetime
(\taus) are fixed to the published average values \cite{pdg98}.  The
regeneration amplitude is floated in the fit, but constrained to have
a power-law dependence on kaon energy, with the phase determined by
analyticity \cite{e731prd,analphi}.
%
%
%
The kaon energy distributions are allowed to be different for the
$\pi^+ \pi^-$ and $\pi^0\pi^0$ modes, with a floating normalization
correction in each energy bin for each mode (24 fit parameters).
Fitting was done ``blind'', by hiding the value of
\reepoe\ with an unknown offset, until after the analysis and
systematic error evaluation were finalized.  The result is
$\reepoe = (28.0 \pm 3.0) \times 10^{-4}$, where the error is
statistical only.  The fit $\chi^2$ is 30 for 21 degrees of freedom.

As a general rule, only biases which affect the \KL\ and \KS\ samples
differently will lead to systematic errors on \reepoe.  Possible
sources may be divided into four classes: (1) data collection
inefficiencies; (2) biases in event reconstruction, sample selection,
and background subtraction; (3) misunderstanding of the detector
acceptance; and (4) uncertainties in kaon flux and physics parameters.
Table~\ref{tb:syst} summarizes all of the estimated contributions;
only those that are large or require special explanation will
be discussed below.

\begin{table}
\centering
\caption{Systematic uncertainties on \reepoe.
}
\begin{tabular}{ldd} 
            & \multicolumn{2}{c}{Uncertainty ($\times 10^{-4})$} \\
 Source of uncertainty &  from $\pi^+\pi^-$  &  from $\pi^0\pi^0$  \\
\hline 
\hline 
\multicolumn{3}{l}{Class 1: Data collection} \\
 \hspace*{0.1in} Trigger and Level~3 filter   & 0.5 & 0.3 \\
\hline 
\multicolumn{3}{l}{Class 2: Event reconstruction, selection, backgrounds} \\
 \hspace*{0.1in} Energy scale & 0.1 & 0.7 \\
 \hspace*{0.1in} Calorimeter nonlinearity & --- & 0.6 \\
 \hspace*{0.1in} Detector calibration, alignment &  0.3 & 0.4 \\
 \hspace*{0.1in} Analysis cut variations & 0.6 & 0.8 \\
 \hspace*{0.1in} Background subtraction & 0.2 & 0.8 \\
\hline 
\multicolumn{3}{l}{Class 3: Detector acceptance} \\
 \hspace*{0.1in} Limiting apertures  & 0.3 & 0.5 \\   
 \hspace*{0.1in} Detector resolution  &  0.4 & $<$0.1 \\
 \hspace*{0.1in} Drift chamber simulation & 0.6 & --- \\
 \hspace*{0.1in} $Z$ dependence of acceptance & 1.6  &  0.7 \\
 \hspace*{0.1in} Monte Carlo statistics & 0.5 &  0.9  \\
\hline 
\multicolumn{3}{l}{Class 4: Kaon flux and physics parameters} \\
 \multicolumn{3}{l}{\hspace*{0.1in} Regenerator-beam attenuation:} \\
   \hspace*{0.25in} 1996 versus 1997 & \multicolumn{2}{d}{ $0.2$} \\
   \hspace*{0.25in} Energy dependence & \multicolumn{2}{d}{ $0.2$} \\
 \hspace*{0.1in} \delm, \taus, regeneration phase  &
                                       \multicolumn{2}{d}{ $0.2$} \\
\hline 
 TOTAL  &                              \multicolumn{2}{d}{ $2.8$} \\
\end{tabular} 
\label{tb:syst}
\end{table}

Two of the largest uncertainties in the second class are related to
the measurement of photon energies by the calorimeter.  A systematic
shift in measured energies can shift the reconstructed $Z$ vertex and
\EK\ distributions for the $\pi^0\pi^0$ sample and thus can bias
\reepoe, mainly by moving \KL\ events past the fiducial $Z$ cut at
158~m.  After calibrating the calorimeter with electrons (and allowing
for a small expected electron-photon difference), a final energy
scale correction for photons of $-0.125$\% is determined by matching
the sharp turn-on of the $\pi^0\pi^0$ $Z$ distribution at the
regenerator edge between data and MC.
After making this correction, a check using $\pi^0$ pairs produced by
hadronic interactions in the vacuum window reveals a $Z$ mismatch of
2~cm at the downstream end of the decay region, leading to a
systematic error of $0.7 \times 10^{-4}$ on \reepoe.
%
%
Residual
nonlinearities in the calorimeter response, studied from the variation
of the mean $\pi^0\pi^0$ invariant mass as a function of \EK,
contribute an additional error of $0.6 \times 10^{-4}$.

We assign systematic errors based on the dependence of the measured
value of \reepoe\ on variations of key analysis cuts, in particular
the \ptsq\ cut for $\pi^+\pi^-$ and the ring-number and photon quality
cuts for $\pi^0\pi^0$.  No significant dependence is observed on other
analysis cuts.

The accuracy of the background determination for the $\pi^0\pi^0$
samples depends on our understanding of kaon scattering in the
regenerator and collimator.  We consider several variations in
scattering models and in the procedures for tuning the MC with 
$K \to \pi^+\pi^-$ decays; these affect the shapes of the background MC
ring-number distributions, but the sideband normalization procedure
limits the impact on \reepoe.  We assign an uncertainty of $0.8 \times
10^{-4}$.
%

\begin{figure}
\centering
\hspace*{0.0in}\psfig{figure=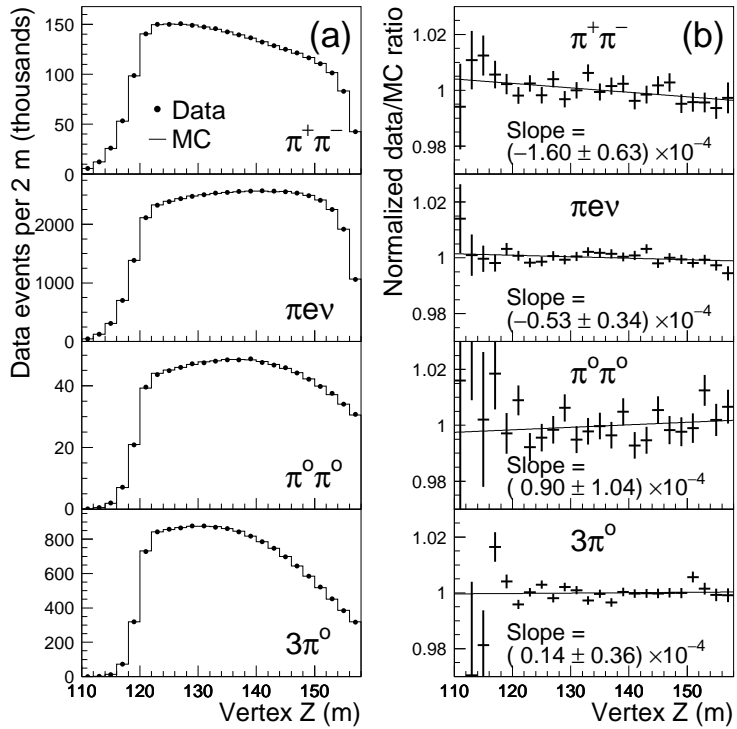,width=3.9in}
\caption{(a) Data versus Monte Carlo comparisons of vacuum-beam $Z$
 distributions for $\pi^+\pi^-$, $\pi e \nu$, $\pi^0\pi^0$, and
 $3\pi^0$ decays.  (b) Linear fits to the data/MC ratio of $Z$
 distributions for each of the four decay modes.}
\label{fig:datamc}
\end{figure}

The third class of systematic uncertainties, related to detector
acceptance, contributes the most to the total systematic error.
Many potential detector modeling problems would affect the acceptance
as a function of $Z$, so a crucial check of our understanding of the
acceptance is to compare the $Z$ distribution for the data against the
MC simulation.  Figure~\ref{fig:datamc}
shows the vacuum-beam comparisons for the $\pi^+\pi^-$ and
$\pi^0\pi^0$ signal modes as well as for the much larger $\pi e \nu$
and $3\pi^0$ samples.
The overall agreement is very good, but since the mean $Z$ positions
for \KL\ and \KS\ decays differ by about 6~m, a relative slope of 
$10^{-4}$ per meter in the data/MC ratio would cause an error of
$10^{-4}$ on \reepoe.  As shown in Fig.~\ref{fig:datamc}~(b), the $\pi
e \nu$ comparison agrees to better than this level; however, the
$\pi^+\pi^-$ comparison has a slope of $(-1.60 \pm 0.63) \times
10^{-4}$ per meter.  Although the statistical significance of the
$\pi^+\pi^-$ slope is marginal, we assign a systematic
error on \reepoe\ based on the full size of the apparent slope, $1.6 \times
10^{-4}$.
The $3\pi^0$ and $\pi^0\pi^0$ $Z$ distributions agree well, and we
place a limit of $0.7 \times 10^{-4}$ on the possible \reepoe\ bias
from the neutral-mode acceptance.

Other checks on the acceptance include data/MC comparisons of track
illuminations at the drift chambers and CsI, photon illumination at
the CsI, and minimum photon separation distance.  These all agree well
and indicate no other sources of acceptance misunderstanding.

The final class of systematic uncertainties includes possible
differences in the \KS/\KL\ flux ratio between the $\pi^+\pi^-$ and
$\pi^0\pi^0$ samples.
The flux ratio is nominally the same for the 1996 and 1997 running
periods because the same regenerator and movable absorber were used;
however, we must assign a small uncertainty on \reepoe\ due to a
possible temperature difference which would change their densities and
thus the regenerator-beam attenuation.
%
%
In addition,
the $\pi^+\pi^-$ and $\pi^0\pi^0$ samples have somewhat different
energy distributions, so the uncertainty in the energy dependence of
the attenuation (measured using $\pi^+\pi^-\pi^0$ and $3\pi^0$ data)
leads to a small uncertainty on \reepoe.

Finally, we assign uncertainties corresponding to one-sigma variations
of \delm\ and \taus\ from the published averages \cite{vardmts}, and from a
deviation of the phase of the regeneration amplitude by $\pm
0.5^\circ$ from the value given by analyticity \cite{analphi}.
Adding all contributions in quadrature, the total systematic
uncertainty on \reepoe\ is $2.8 \times 10^{-4}$.

We have performed several cross-checks on the \reepoe\ result.
Consistent values are obtained at all kaon energies, and there is no
significant variation as a function of time or beam intensity.
Relaxing the power-law constraint on the regeneration amplitude yields
a consistent value with the same precision.
%
%
%
We have also extracted \reepoe\ using an alternative fitting technique
which compares the vacuum- and regenerator-beam $Z$ distributions
directly, eliminating the need for a Monte Carlo simulation to
determine the acceptance.  While less statistically powerful, this
technique yields a value of \reepoe\ which is consistent
with the standard analysis based on the uncorrelated parts of the
statistical and systematic errors.
Finally, using $\pi^+\pi^-$ data from 1996 (collected simultaneously
with the $\pi^0\pi^0$ data) instead of from 1997 yields a value of
\reepoe\ which is consistent with the standard analysis, allowing a
systematic error of $4 \times 10^{-4}$ due to the 1996 Level~3
inefficiency.

In conclusion, we have measured \reepoe\ to be $(28.0 \pm
3.0~\text{(stat)} \pm 2.8~\text{(syst)}) \times 10^{-4}$; combining
the errors in quadrature, $\reepoe = (28.0 \pm 4.1) \times 10^{-4}$.
This result definitively establishes the existence of \CP\ violation
in a decay process, agreeing better with the earlier measurement from
NA31 than with E731 \cite{diffnote}, and
shows that a superweak interaction cannot be the sole source of
\CP\ violation in the $K$ meson system.
The average of the three measurements, $(21.7 \pm 3.0) \times
10^{-4}$, while at the high end of standard-model predictions,
supports the notion of a nonzero phase in the
CKM matrix.  Further theoretical and experimental advances are needed
before one can say whether or not there are other sources of \CP\
violation.


We gratefully acknowledge the support and effort of the Fermilab
staff and the technical staffs of the participating institutions for
their vital contributions.  This work was supported in part by the U.~S. 
Department of Energy, The National Science Foundation, and The Ministry of
Education and Science of Japan. 



\begin{references}
%
\bibitem[*]{monnier_on_leave}
  On leave from C.~P.~P. Marseille/C.~N.~R.~S., France.
%
\bibitem{ccft} J. H. Christenson, J. W. Cronin, V. L. Fitch, and
  R. Turlay, Phys.\ Rev.\ Lett.\ {\bf 13}, 138 (1964).
\bibitem{ww_review} For a review, see B. Winstein and L. Wolfenstein,
  Rev.\ Mod.\ Phys.\ {\bf 65}, 1113 (1993).
\bibitem{ckm} M. Kobayashi and T. Maskawa,
  Prog.\ Theor.\ Phys.\ {\bf 49}, 652 (1973).
\bibitem{early_calc} J. Ellis, M. K. Gaillard, and D. V. Nanopoulos,
  Nucl.\ Phys.\ {\bf B109}, 213 (1976);
  F. J. Gilman and M. B. Wise, Phys.\ Lett.\ {\bf B83}, 83 (1979).
\bibitem{ciuchini} M. Ciuchini, Nucl.\ Phys.\ Proc.\ Suppl.\ {\bf 59},
  149 (1997); hep-ph/9701278.
\bibitem{buras98} A. J. Buras, in {\it Probing the Standard
  Model of Particle Interactions} (Elsevier, 1999), edited by R. Gupta
  {\it et al.}; hep-ph/9806471.  This estimate assumes that the mass
  of the strange quark is $m_s(m_c) = 125 \pm 20$~MeV$/c^2$.
\bibitem{bertolini} S. Bertolini {\it et al.}, Nucl.\ Phys.\
  {\bf B514}, 93 (1998); hep-ph/9706260.
\bibitem{superweak} L. Wolfenstein, Phys.\ Rev.\ Lett.\ {\bf 13}, 562
  (1964); Comments Nucl.\ Part.\ Phys.\ {\bf 21}, 275 (1994).
\bibitem{e731} L. K. Gibbons {\it et al.},
  Phys.\ Rev.\ Lett.\ {\bf 70}, 1203 (1993).
\bibitem{na31} G. D. Barr {\it et al.}, Phys.\ Lett.\ {\bf B317}, 233 (1993).
\bibitem{e731prd} L. K. Gibbons {\it et al.},
  Phys.\ Rev.\ D {\bf 55}, 6625 (1997).
\bibitem{beam} V. Bocean {\it et al.},
  Fermilab Report TM-2046, 1998 (unpublished).
%
\bibitem{na31train} In contrast, NA31 used a movable \KS\ target to
make the $Z$ distributions similar for \KS\ and \KL\ decays, so that
the acceptance nearly canceled in the \KS/\KL\ ratio.
%
\bibitem{csical} A. Roodman, in {\it Proceedings of the Seventh
  International Conference on Calorimetry in High Energy Physics},
  edited by E. Cheu {\it et al.} (World Scientific, 1998), p.\ 89.
\bibitem{dpmt} J. Whitmore,
  Nucl.\ Instrum.\ Methods Phys.\ Res., Sect.\ A {\bf 409}, 687 (1998).
%
\bibitem{deadtime} These veto elements, primarily the regenerator,
cause a deadtime of 12\% (identical for decays in either beam).
%
\bibitem{hcc} C. Bown {\it et al.},
  Nucl.\ Instrum.\ Methods Phys.\ Res., Sect.\ A {\bf 369}, 248 (1996).
\bibitem{geant} R. Brun {\it et al.}, computer code {\sc GEANT} 3.21,
  CERN, Geneva, 1994.
  Hadronic interactions are simulated using the {\sc FLUKA} package.
\bibitem{pdg98} Particle Data Group, C. Caso {\it et al.},
  Eur.\ Phys.\ J. C {\bf 3}, 1 (1998).
\bibitem{analphi} See R. A. Briere and B. Winstein, Phys.\ Rev.\ Lett.\
  {\bf 75}, 402 (1995).
%
\bibitem{vardmts} If \delm\ increases by $0.0014 \times
10^{10}$~$\hbar$~$s^{-1}$, then \reepoe\ changes by $+0.05 \times
10^{-4}$.  If \taus\ increases by $0.0008 \times 10^{-10}$~$s$, then
\reepoe\ changes by $-0.11 \times 10^{-4}$.
%
KTeV will publish new high-precision measurements of \delm\ and \taus\
in the future.  Preliminary fits give values which are consistent
between the $\pi^+\pi^-$ and $\pi^0\pi^0$ data samples.
%
\bibitem{diffnote} Scrutiny of E731 has not revealed any explanation
for its lower measured value other than a possible, if improbable,
fluctuation.
\end{references}
\end{document}